# A revised framework for the assessment of psychological safety in autonomous vehicles

Yandika Sirgabsou[a*], Benjamin Hardin[b], François Leblanc[a], Efi Raili[c], David Jackson[c], Pericle Salvini[b], Lars Kunze[d], Marina Jirotka[b]

[a] *Capgemini Engineering, Toulouse, France;*

[b] *Dept. of Computer Science, University of Oxford, UK;*

[c] *Capgemini Engineering, UK;*

[d] *Bristol Robotics Laboratory, University of the West of England, UK;*

*Yandika Sirgabsou @ yandika.sirgabsou@capgemini.com; Capgemini Engineering, 4 Av. Didier Daurat, 31700 Blagnac, France

**ABSTRACT**

Despite recent technological progress in the development of autonomous vehicles (AVs), their societal acceptability remains a subject of debate as recent research findings point to psychological roadblocks. Concerns arise not only for physical safety but also for potential psychological risks resulting from human interaction with AVs. Psychological concepts such as trust, and perceived safety are well-studied in this context and are found to be determinant factors for the intention to use AVs. Unfortunately, there has been no formalization of the mechanism by which human interaction with AVs may lead to psychological hazards, threatening trust, perceived safety, and acceptability. Furthermore, there has been little prior research that conceptualizes the severity of psychological risk in AVs, and there are no clear guidelines for a systems designer on how to assess and address psychological risk in the AV development context. Having a psychological severity scale would enable developers, testers, and regulators to objectively assess and mitigate psychological risk of human-AV interaction. To address these limitations, this paper extends a theoretical framework for AV psychological safety risk assessment based on an early proposal. The proposed framework consists of an extended risk model for psychological safety including all the key concepts related to psychological safety in AVs, and an assessment method based on the Systems-Theoretic Accident Model and Processes (STAMP). We demonstrate the usefulness of the theoretical framework through a highly automated AV use case scenario, uncovering factors which may lead to psychological risk for an occupant. The use cases provide examples of how to use the framework to extensively evaluate psychological risk and determine vehicle behaviour that could lead to this risk. By developing a theoretical framework for AV psychological safety risk assessment, we provide a foundation and a method that were previously lacking to better enable responsible AV development regarding psychological safety.

**Keywords:** psychological safety, autonomous vehicles, systems safety, trust, acceptability

## 1. INTRODUCTION

The advent of autonomous vehicles (AVs) represents a significant leap in transportation technology, promising enhanced safety and efficiency on our roads. However, despite these technological advancements, the widespread acceptance of AVs faces considerable challenges. Recent research has uncovered psychological barriers that extend beyond concerns for physical safety, highlighting potential psychological risks associated with human-AV interactions (Shariff et al., 2017). While concepts such as trust and perceived safety have been extensively studied in the context of AVs and are recognized as influencing factors of intention to use these vehicles, there remains a significant gap in our understanding, especially how to address them from a systems safety approach. Specifically, there has been a lack of formalization regarding the mechanisms through which human interaction with AVs may lead to psychological hazards, potentially undermining trust, perceived safety, and overall acceptability. Furthermore, although a number of psychological risk scales related to these hazards have been proposed in the fields of psychology or occupational health (Taibi et al., 2022; Winwood et al., 2012), there has been limited research on conceptualizing the severity of psychological risk in AVs, leaving systems designers without clear guidelines for assessing and addressing these risks during the development process.

This paper builds upon an earlier proposal to extend a theoretical framework for AV psychological safety risk assessment (Sirgabsou et al., 2024). This framework incorporates an expanded risk model for psychological safety, encompassing key concepts related to psychological safety in AVs, and proposes an assessment method based on the Systems-Theoretic Accident Model and Processes (STAMP) approach. By applying this framework to one highly automated AV use case scenario, we demonstrate its practical utility in evaluating psychological risk and identifying vehicle behaviours that could lead to such risks.

The remainder of the paper is structured as follows. The next section discusses the state of the art, noting a few limitations in the field. The third section describes the method used in the development of the proposed framework. The fourth section outlines the proposed extended psychological safety risk model and its adaptation to the systems-theoretic approach for psychological hazard analysis. The fifth section is an application of the proposal to a use case scenario. Section six discusses the added values and limitations of the proposals as well as avenues for future research. Finally, section 7 concludes the paper by summarizing the key points and reemphasizing the need for future works on the subject.

## 2. STATE OF THE ART

To date, safety research for AVs has predominately focused on physical safety aspects. When psychological aspects are discussed, it is primarily in terms of perceived safety, or how safe the occupant views a particular situation in regard to environmental factors and the vehicle’s ability to manage them (Rubagotti et al., 2022). However, we argue that thinking about perceived safety overlooks many of the components of psychological safety that are necessary for ensuring that AVs avoid causing psychological discomfort. Furthermore, a risk assessment method is necessary so that psychological safety concerns can be systematically identified. Below we will outline the current state of the art in physical safety research, discuss the foundations of AV psychological safety, and connect these concepts to develop foundations of a risk assessment method for AV psychological safety.

## 2.1. Psychological safety

### 2.1.1. Definition of psychological safety

The notion of psychological safety was first defined in the occupational context, that is, in work related settings. (See Abror & Patrisia, 2020; Edmondson & Lei, 2014). According to (Edmondson & Lei, 2014), psychological safety is "people's perceptions of the consequences of taking interpersonal risks in a particular context such as a workplace". In this context, psychological safety is understood as the willingness of team members to place themselves in vulnerable positions by being honest about a situation and coming forward without fear of retaliation. But the notion of psychological safety is not limited to the interpersonal risk context. For instance, in collaborative settings, where human agents interact with automated agents such as robots or automated vehicles, psychological safety is a relevant concern that needs to be addressed (Su et al., 2024; Xing et al., 2021; You et al., 2018). For instance, in the field of physical Human-Robot Interaction (pHRI), psychological safety was described as "ensuring that the human perceives interaction with robot as being safe, and that interaction does not lead to any psychological discomfort or stress as a result of the robot's motion, appearance, embodiment, gaze, speech, posture, social conduct, or any other attribute" (Lasota et al., 2017).

### 2.1.2. Psychological safety and autonomous vehicles

Autonomous Vehicles (AVs) can be considered a particular type of robot (Lasota et al., 2017). Consequently, the issue and concepts related to psychological safety found in pHRI (outlined below), are relevant in the AV context. Moreover, perceived safety (a concept synonymous to psychological safety) has been extensively discussed in the AV technology acceptance literature (Cao et al., 2021; Moody et al., 2020; Nordhoff et al., 2021; Othman et al., 2023; Rubagotti et al., 2022). In this light, perceived safety has been highlighted as a reliable predictor of individuals' intentions to use AVs and thus is viewed as a determinant factor for the widespread adoption of AVs (Prasetio & Nurliyana, 2023).

### 2.1.3. Psychological safety components: definition and evaluation methods

In our previous work, we defined psychological safety of AVs as including trust, perceived control, perceived risk, psychological empowerment, responsibility and liability, comfort, predictability, perceived support, effort-reward, and low demands (Sirgabsou et al., 2024). However, these components must be measured in order to estimate their risk. Once their risk has been estimated, efforts for mitigation can be identified based on the most problematic components, and risk analysis can be rerun to estimate the success of design choices in mitigating risk. These components are based on the existing literature for psychological-related concerns of AVs, and here we consolidate them into four major components that underpin the psychological experience of interacting with an AV. These four components include *perceived control, predictability, trust,* and *perceived support*, and some of their largest subcomponents are outlined below. How these four components affect perceptions about AVs, user experience in AVs, and intention to use AVs has been studied extensively and relevant literature is included in the table.

**Table 1. Psychological safety components literature**

| | Component | Subcomponent | Literature |
|---|---|---|---|
| **Psychological Safety** | Trust | General | Jing et al., 2020 |
| | | Situational | Basu & Singhal, 2016 |

| | Component | Subcomponent | Literature |
|---|---|---|---|
| | | Perceived Risk | Zhang et al., 2021 |
| | | Appearance | Baker et al., 2018 |
| | | Swift vs. Traditional | |
| | | Attitudinal vs. Behavioural | |
| | Perceived Control | Usability | Hegner et al., 2019 |
| | | Workload/Demands | Cheng et al., 2019 |
| | | Explanations | Buckley et al., 2018 |
| | | Responsibility/Liability | Kim et al., 2023 |
| | | | Cao et al., 2021 |
| | Predictability | Mental Model of Automation | Cao et al., 2021 |
| | | Familiarity with Automation | Ayoub et al., 2021 |
| | | Familiarity with Environment | Xing et al., 2021 |
| | | Surprise/Shock | Noneka et al., 2004 |
| | Perceived Support | Explanations | Omeiza et al., 2021 |
| | | Accident/Event support | Petersen et al., 2019 |
| | | Feedback, warnings | Eyben et al., 2010 |
| | | Situation awareness support | Lee et al., 2023 |

These four components have existing standards of measurement and evaluation that can be used to estimate risk and build our overall estimation of psychological risk. If the user of our framework desires a more extensive analysis, we further divide these four components into subcomponents and their accompanying evaluation methods. The set measurement method for these psychological safety components and their subcomponents, divided into qualitative and quantitative techniques, is provided in annex. While the qualitative method comprised several scales and self-reported questionnaires, the quantitative ones mostly rely on the measures of physiological variables such as eye gaze, heart rate or heart rate variability.

### 2.1.4. Psychological risk mitigation

To mitigate psychological risks, it is necessary to determine criteria and principles that will be relied on and applied to set psychological safety goals. For this purpose, the following criteria from the literature are identified as useful in preventing or mitigating psychological risks.

**Transparency.** Transparency affects trust. Lee & See, 2004 revealed that appropriate trust and reliance on automation depend on how well the capabilities of the automation are conveyed to the user, and the operator's understanding of how the context affects the capability of the automation. Consequently, Lee & See, 2004 further argued that improving trust can be done by making the algorithms of the automation simpler, revealing their operation more clearly and, by relating the context to the autonomy's capability.

**Feedback.** Feedback is key in addressing some issues related to all the psychological components. For example, feedback has been shown to be a criterion that enhances Situation Awareness (Norman, 1990). In relation to trust, (Stanton & Young, 2000) argued that "the level of trust a human agent may have in an automated system may depend upon appropriate feedback". Therefore, (Stanton & Young, 2000) argued that the development of the driver's trust in the automated system may depend upon appropriate feedback. Feedback can be improved through adequate HMI design considerations.

## 2.2. Physical Safety

### 2.2.1. Classical safety

Safety, in the engineering and physical sense is defined as "freedom from unacceptable risks" where risk combines severity (magnitude of an undesired event) with its likelihood (mathematical expression the probability of occurrence) (*MIL-STD-882 D SYSTEM SAFETY*, 2000). In other words, risk = f (severity, likelihood). System safety is an engineering discipline in its own right, regulated and governed by standards. It applies specialized scientific and engineering principles, criteria, and techniques to identify and eliminate hazards, in order to reduce mishap risk. Such principles include the development of critical systems in accordance with a defined lifecycle process, where well-specified safety related activities such HARA (Hazard and Risk Analysis) must be conducted in order to specify measures that must have certain defined criteria in terms of acceptability, completeness, and traceability criteria.

In the automotive industry, the level of rigor to observe in order to achieve acceptable risk in the development of an item is dependent on its Automotive Safety Integrity Level (ASIL) rating depending on the level of risk. Depending on the ASIL level ranging from A (the least constringent) to D (the most constringent), the ISO 26262 standard (*ISO 26262 [2018] Road Vehicles-Functional Safety-Part 1*, 2018) recommends or requires certain methods and quality management principles to be applied in the development of systems, hardware or software components. As for safety analysis, a combination of inductive or bottom-up methods that start from known causes and identify possible effects, and deductive or top-down methods that start from known effects and seek possible causes are recommended or required based on ASIL levels. Inductive analyses and deductive analyses complement each other and therefore increase the coverage of their result. Furthermore, the analysis can be quantitative or qualitative. Quantitative analysis methods predict the frequency of failures whereas qualitative analysis methods identify failures but do not predict the frequency of failures.

### 2.2.2. Systems theoretic approach to safety

STAMP (System-Theoretic Accident Model and Processes) is an accident causation model based on system and control theory (N. G. Leveson, 2012). In system theory, safety is viewed as an emerging property arising from complex relationships among system parts and is treated as a "dynamic control problem" rather than a component failure problem. STAMP defines safety as "freedom from unacceptable losses (as determined by the system stakeholders)". To analyse safety, STAMP relies on the Systems-Theoretic

Process Analysis (STPA) that utilizes an abstract model called hierarchical safety control (N. G. Leveson, 2017).

### 2.3. Problem statement

Despite the recognised importance of psychological safety for AVs adoption, it is not well understood, nor addressed in the system safety context. In our previous work (Sirgabsou et al., 2024), we attempted to build on the system safety context and address psychological safety in AVs through the proposal of a risk model for psychological safety of AVs and an analysis method based on STAMP and STPA. However, our previous work did not consider how to measure components of psychological safety at a basic level, and we did not specify the principles on which were based the control measures used in our psychological safety goals. However, this measurement is critical, as it creates the foundation for the rest of the risk analysis to build on, and the mitigation measures proposed for psychological safety must be based on well-established principles and criteria. Therefore, our next ambition is to extend our proposed risk model and further improve our adaptation of the systems theoretic hazard analysis by ensuring complete traceability between the analysis artefacts and identifying areas to mitigate based on the quantification of the psychological risk relying on the measurement of the specified psychological components.

## 3. METHOD

Our strategy to develop this framework relied on an analytic review of existing options to address psychological safety. Subsequently, a literature on human factors analysis (Energy Institute, 2020; NASA Chang, 2006), psychological safety, and physical safety was conducted to understand the very essence of psychological safety and the adequation of the state-of-the-art analysis techniques for its assessment.

In this approach, we first made a state of the art on the very notion of psychological safety in several fields (including those of well-being at work and physical interaction between humans and robots) to understand its underlying principles. Psychological safety is complex, suggestive, having more to do with perceptions and impressions and not necessarily justified by the objective and real aspect of the real risk (Lasota et al., 2017). Thus, behaviour that is completely safe but unusual from a physical point of view of the autonomy, and whose intention is not well communicated to the human, can be perceived as dangerous or even a threat.

Secondly, we did a state-of-the-art study of physical safety, including both the classical and emerging techniques of the industry, to assess whether they were adequate for addressing psychological safety. From this study, it was found that from the point of view of classical safety techniques, physical safety was deterministic and objective in its definition based on the notion of acceptability of risk expressed mathematically as a product of severity and probability (DoD, 2012; *ISO 26262 [2018] Road Vehicles-Functional Safety-Part 1*, 2018). We then looked at the more recent safety approaches, the main ones being the SotIF (*ISO 21448:2022 - Road Vehicles — Safety of the Intended Functionality*, n.d.) and the UL4600 (Koopman et al., 2019) and the systems-theoretic approach (N. G. Leveson, 2012; N. Leveson & Thomas, 2018). The first two are recent standards aimed at addressing safety issues related to autonomy, in the automotive field. However, their focus and the methods they offer remain limited to the physical aspects of safety. In the systems-theoretic approach, we found an adequate method for addressing psychological safety in the sense that it is not limited to the physical aspects of risk (N. Leveson, 2020). Instead, it makes provisions for addressing other types of risks that are not necessarily physical (such as financial, reputational, or psychological…) defined as losses deriving from the violation of stakeholders needs. Hence, beyond random and systematic failures already addressed by existing standards, the theoretical-system approach allows us to methodically and more effectively address other emerging phenomena such as psychological safety that may occur in the context of interactions in complex systems.

From these preliminary studies, we deduced that although psychological and physical safety are linked, they are governed by different mechanisms. This allowed us to conclude on the inadequacy of conventional methods to address physical safety. Subsequently, a literature review was conducted on measurements and evaluation techniques for the identified key psychological safety components. This literature review revealed that these measurement techniques can be relied upon for the determination of psychological risk.

# 4. AN EXTENDED PSYCHOLOGICAL SAFETY RISK MODEL AND HAZARD ANALYSIS METHOD FOR AVS

## 4.1. Extended psychological safety risk model for autonomous vehicles based on the System Theoretic Accident Model and Processes (STAMP)

The proposed psychological risk model extends the previous model by refining the psychological state into the previously described four psychological state components (trust, perceived control, predictability, and perceived support). See Figure 1. It is hypothesized that objective measures (e.g. anxiety, trust, perceived control...) of the different psychological components can enable a tangible determination of the psychological state, enabling the determination of the severity used in the PsySIL parameter. Therefore, this refinement aims to allow a pragmatic estimation of the severity of the psychological losses and assignment of PsySIL based on the measure of each component.

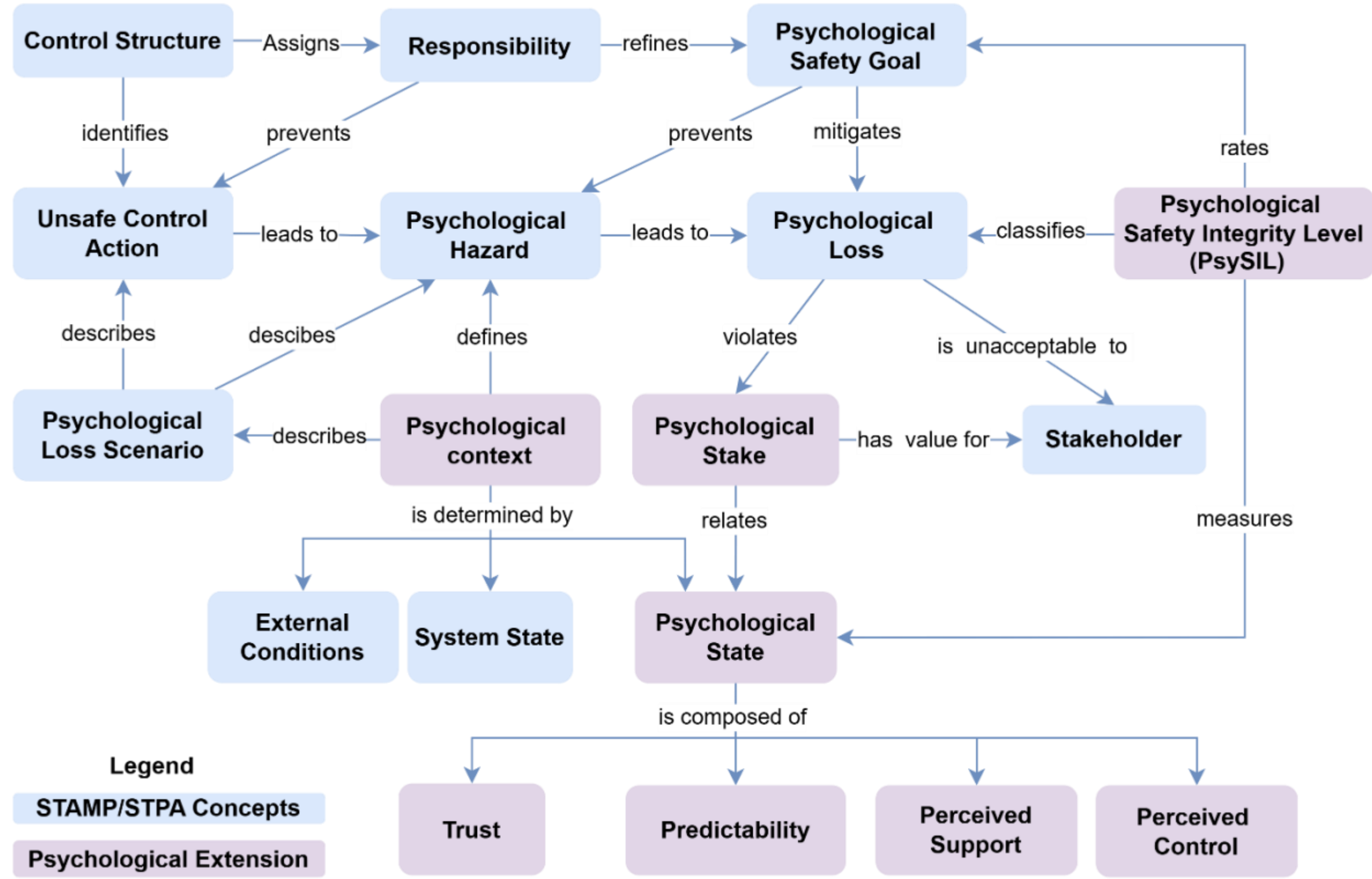

Figure 1. Extended psychological safety risk model for AVs

Additionally, the extended model adds traceability relations to the STPA artefacts (including responsibility, unsafe control actions and psychological loss scenarios). Developing the extended model enables us to ensure that relevant elements of the risk model are methodically covered by the analysis. A psychological loss scenario explains how a particular operational and psychological context will lead to an Unsafe Control Action (UCA), psychological hazard & loss. Responsibilities are declinations or assignments of psychological safety goals into functional safety requirements (what each component in the control structure must do to enforce a particular psychological safety goals).

Finally, the new risk model introduces an element of psychological context which is key in both defining psychological hazards and describing psychological loss scenario (i.e. how psychological losses occur). Based on our definition of a psychological hazard, this psychological context is composed of three components: the internal vehicle state, external conditions or events, and the psychological state of the human driver as illustrated in Figure 1.

### 4.2. Psychological hazard analysis & risk assessment method based on the systems-theoretic process analysis

Based on the proposed risk model, this methodology adapts the 4 steps of STPA to capture the psychological dimension of risk as illustrated in Figure 2.

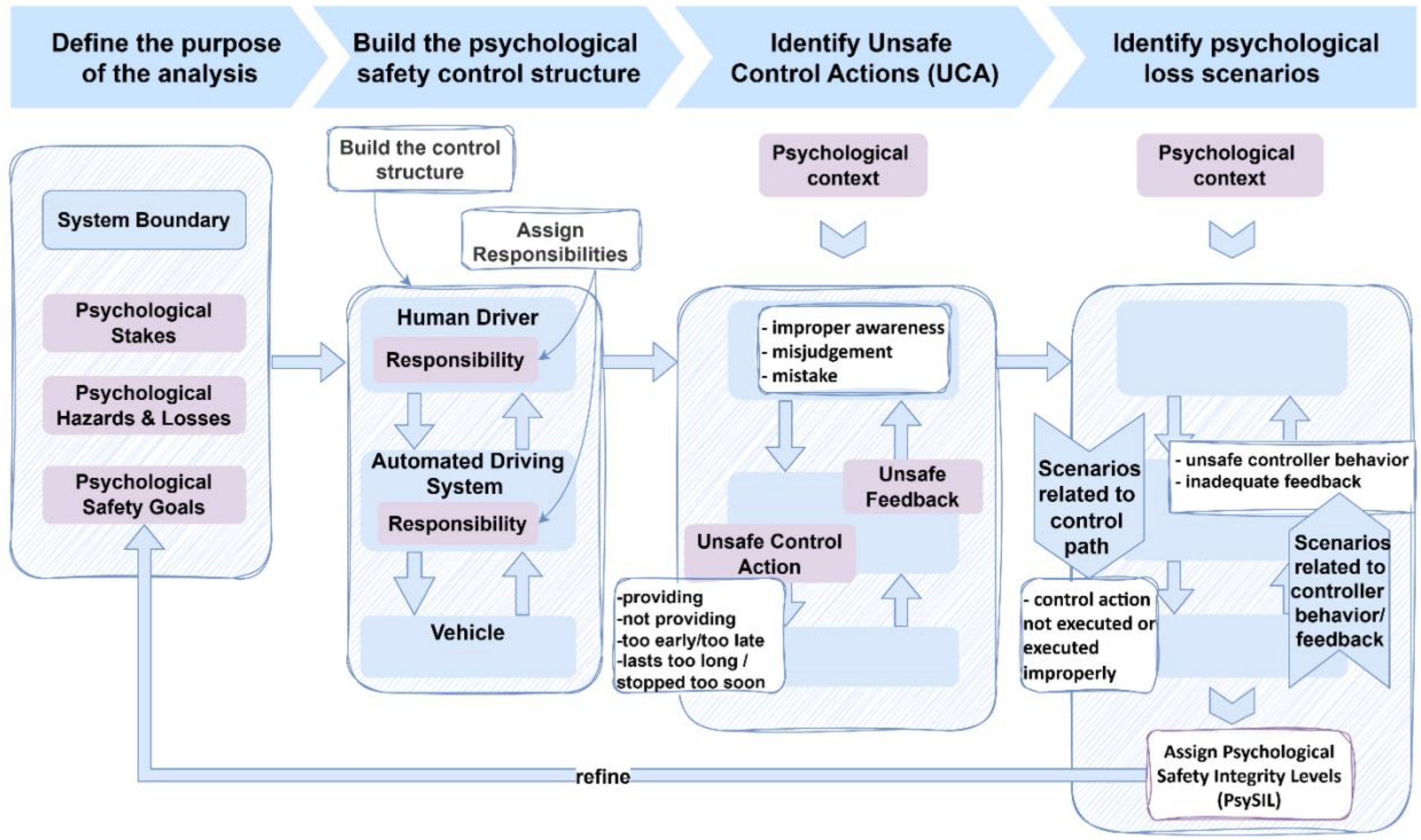

**Figure 2. Psychological systems-theoretic process analysis.**

The first step of the methodology defines the purpose of the analysis (which is to perform preliminary psychological hazards and risk assessment). In this step, the system boundary is defined, psychological losses (harm) and hazards identified, and psychological safety goals are defined as preventive or mitigative measures.

The second step builds the psychological safety control structure that will be used throughout the analysis. This step also assigns responsibilities to components of the control structure to ensure the enforcement of the previously specified psychological safety goals.

The third step identifies psychologically Unsafe Control Actions (UCA). As defined in the risk model, UCAs are potential cause of hazards (psychological in this case) occurring in a defined context (e.g. during a particular operational phase). STAMP defines four categories of UCA related to control actions not being provided or provided inadequately. They include: 1) not providing the control action, 2) providing the control action (for instance unexpectedly, in excess, or too quickly), 3) providing a potentially safe control action but too early, too late, or in the wrong order, and finally 4) providing a control action that lasts too long or is stopped too soon. To identify the cause of psychological hazards, this step considers a control action, checks which of the modes lead the psychological hazards identified in the first step of the methodology using the suggested four generic categories UCAs suggested by STAMP. Note that given a psychological hazard, the four categories are not always relevant. However, checking a control action against the four categories ensures that control actions are checked for all potential causes of psychological hazard.

Finally, the fourth step identifies psychological loss scenarios. In this adaptation, the step completes the analysis with a risk assessment which assign PsySIL to each loss scenario and refines the initially specified psychological safety goals.

## 5. USE CASE APPLICATION

The objective of the use case is to apply the extended risk model an adapted STPA analysis methodology to a pedagogical example in order to validate the proposal.

### 5.1. System of interest

The system of interest consists of an Autonomous Vehicle under design. Vehicle occupants are considered part of the system while external factors such as other road users or other objects are considered part of the environment. In terms of autonomy, the use case focuses on SAE level 4 also known as "High Automation", as defined by the Society of Automotive Engineers (SAE MOBILUS, 2021). According to SAE's taxonomy, a level 4 autonomous vehicle is capable of handling all driving tasks under specific conditions without requiring human intervention including monitoring the driving environment and responding to unexpected events and failures. Unlike in lower levels of automation (L1-L3), the human driver is not required to take control but has the option to do so. If the vehicle encounters a scenario it cannot handle, it must achieve a Minimal Risk Condition (MRC) such as safely pulling over or stopping in path in severe cases. Nevertheless, a level 4 AV is still limited to specific Operational Design Domains (ODD) unlike Level 5, which operates in all conditions. ODD which refers to the specific conditions under which an Automated Driving System (ADS) is designed to function is crucial for understanding the boundaries of the systems capabilities, test cases, and other analyses that rely on critical scenario identification.

#### 5.1.1. System architecture

For architectural considerations, the use case relies on SAE J3016 (SAE MOBILUS, 2021), which provides a standardized functional model describing automated driving features. SAE J3016 divides driving automation features into three classes of functions, including strategic, tactical, and operational functions as shown in Figure 3. Strategic Functions involve high-level planning, including setting the destination and determining the general route to be taken. Tactical Functions encompasses Objects and Event Detection and Response (OEDR). It involves the system's ability to detect and appropriately respond to various objects and events encountered during driving. Operational Functions relate to the basic control of the vehicle's motion. This includes both lateral control (steering) and longitudinal control (acceleration and braking).

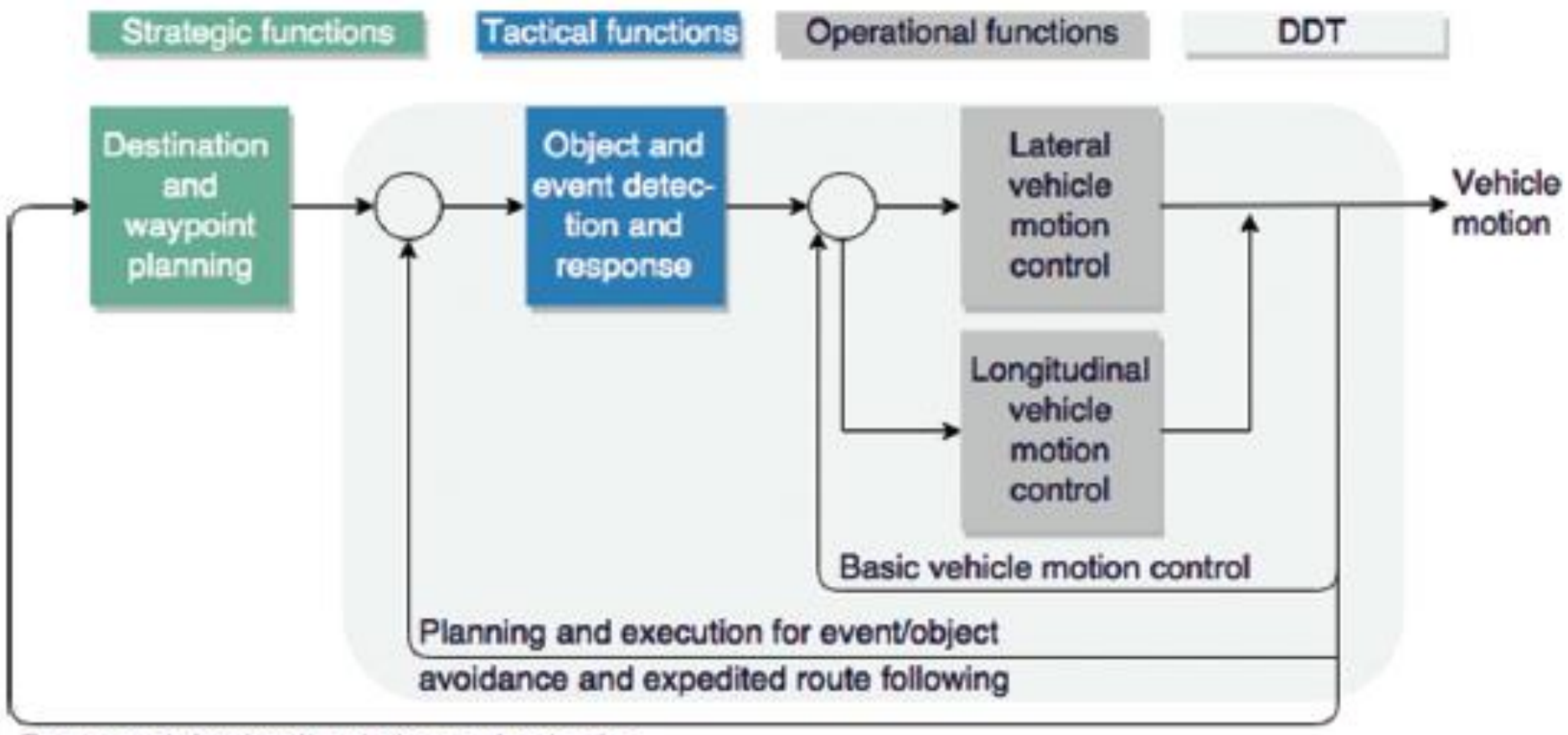

Figure 3. Automated driving functional components classification (SAE MOBILUS, 2021; Serban et al., 2020)

### 5.1.2. Example scenario

As specified in the previous section 0, the third and fourth steps of the methodology require a defined context. This use case suggests using variables derived from the sets of the ODD attributes for defining such context. In general, an ODD is defined by 3 categories of attributes: dynamic elements (movable objects and actors with which the automated vehicle interacts), scenery elements (non-movable element of the vehicle's operating environment) and environmental conditions (the weather or another atmospheric condition including connectivity data). However, it is acknowledged if all possible scenarios were to be considered, the resulting number of scenarios would be extremely high. Therefore, this use case implemented a single illustrative scenario and recommends using existing state of the art tools, methods and criteria to ensure completeness (which is beyond the scope of this use case). For instance, for critical scenario selection, it is suggested that one could choose to focus on edge case scenarios (i.e. approaching ODD limits) and scenarios with the potential worst outcomes (in the psychological sense in this case). For this use case, the following inner-city example scenario is constructed.

- Dynamic elements: ego vehicle operation = vehicle is making a turn, other vehicles = following, oncoming left, pedestrian = pedestrian and dog crossing.
- Scenery elements: objects off-roadway = sidewalk parked cars.
- Environmental conditions: climate = visibility = low, time of day = dawn, shape of road = perpendicular, road conditions = wet.

**Scenario description.** The ego vehicle (white vehicle in the Figure 4) makes a tight turn onto a perpendicular street with limited visibility due to vegetation and many parked cars on both sides of the street. As the vehicle is approaching the turn, the occupant notices some movement behind a parked car on the street that the vehicle is turning onto. As the vehicle makes the turn, a pedestrian pulled by a dog emerges from between parked cars and steps into the street. The pedestrian and dog notice the vehicle and stop short, only partially into the street. However, instead of slowing down and navigating around the pedestrian and dog, the AV makes an abrupt emergency stop assuming collision with the new actor that has appeared in the environment. In this way, the AV has escalated a simple and comfortable evasive manoeuvre into an abrupt emergency stop that surprises the car occupant and the dog walker. The occupant is worried about vehicles that might rear end the stopped AV due to abrupt stop, and its stop in a position of limited visibility, having just made the turn beside parked cars.

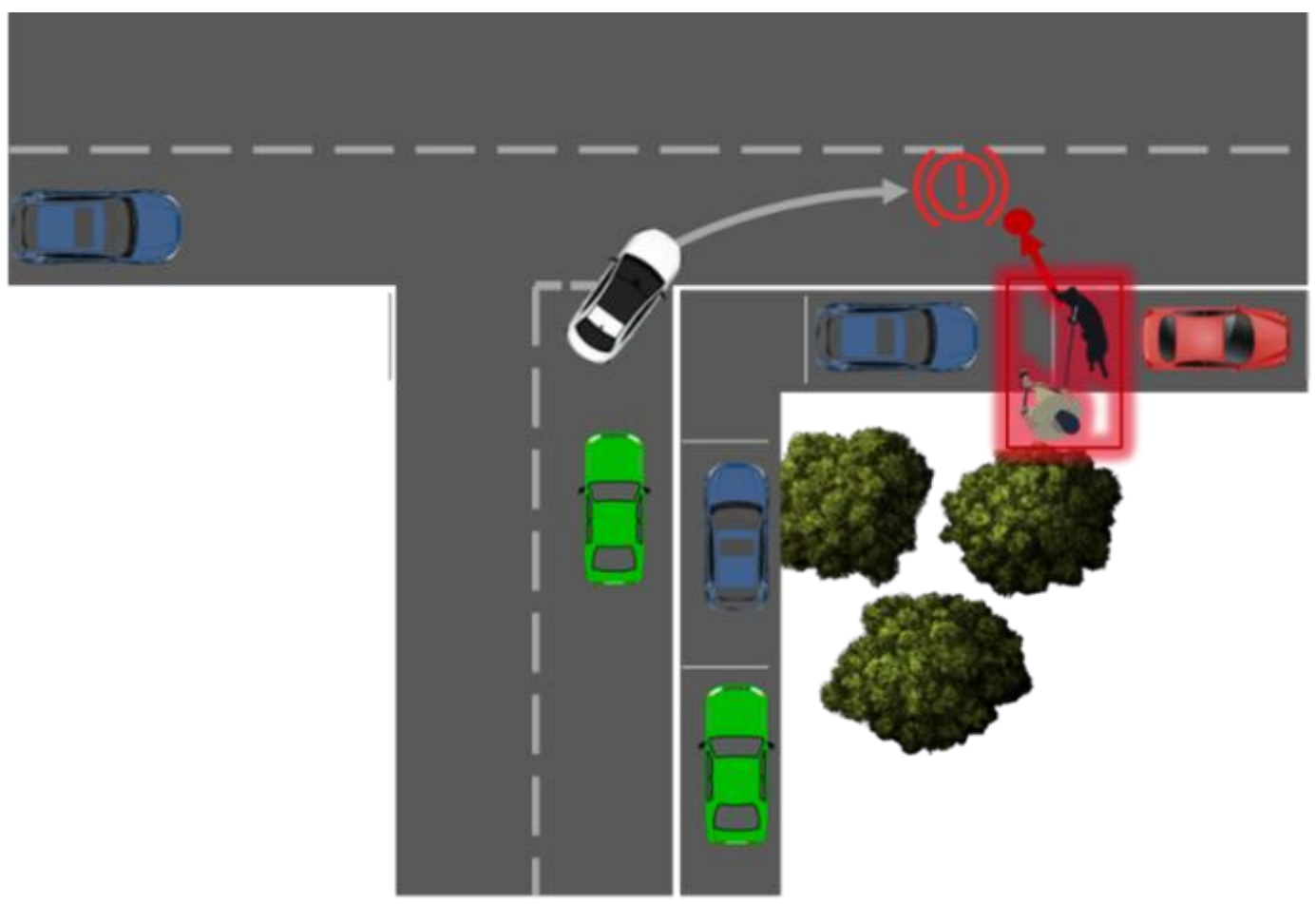

Figure 4. An autonomous vehicle (white), turning onto a street with limited visibility, abruptly stops when a pedestrian and dog partially emerge from between parked cars

### 5.2. Use case application of the risk model and system-theoretic hazard analysis

Having described the system, the purpose of the analysis is to apply the concepts defined in the risk model to and to perform psychological safety assessment involving the operation of an autonomous vehicle.

#### 5.2.1. Define the purpose of the analysis

Defining the purpose of the analysis is done through:

a) the identification of psychological losses and hazards
b) the definition of systems boundary (already previously defined)
c) the specification of psychological safety goals

#### Identify psychological losses

The identification of psychological losses derives from the violation of the considered stakeholder stakes. In this case, these stakes are constituted by the four psychological safety components described in Section 3 including include trust, perceived control, predictability, and perceived support. Relying on the stakes derived from these components and subcomponents, we identified the potential psychological losses listed in Table 2.

Table 2. Identification of psychological losses derived from psychological stakes related to the four psychological state components and subcomponents including - trust, perceived control predictability and perceived support.

| Psychological stake | ID | Psychological loss | Severity |
|---|---|---|---|
| **Trust / perceived safety** | **L1** | **Loss of trust/confidence in the autonomous vehicle** | S3 |
| | L1.1 | Apprehension | S2 |

| | L2 | Loss of perceived safety | S3 |
|---|---|---|---|
| | L1.1 | Stress | S1 |
| | L2.2 | Fear/anxiety | S3 |
| **Perceived Control** | **L3** | **Loss of perceived control** | S2 |
| | L3.1 | Fear/anxiety of not being in control | S3 |
| **Predictability** | **L4** | **Loss of predictability** | S3 |
| | L4.1 | Surprise | S1 |
| | L4.2 | Shock | S3 |
| | L4.3 | Fear/anxiety | S2 |
| **Perceived support** | **L5** | **Loss of perceived support** | S1 |
| | L5.1 | Loss of explainability | S1 |

For each loss, a severity rating is assigned based on (Taibi et al., 2022). While this rating will not be used in this step of the methodology, it provides an early approximation of the gravity of each potential psychological loss. While this analysis focuses on the four psychological safety components outlined above, it is acknowledged that other stakes may be considered, depending on the context or focus of the analysis.

### Identify psychological hazards

Subsequently to the identification of psychological losses, the following psychological hazards listed in Table 3 were identified from the literature mentioned previously. Each hazard is identified by an ID and related to one of more psychological losses.

Table 3. Identified psychological hazard and related losses

| ID | Psychological hazards | Related losses |
|---|---|---|
| H1 | Lack of familiarity with automation or the environment by human driver | L1, L2, L5 |
| H2 | Lack of proper awareness of operational situation by human driver | L1, L2, L5 |
| H3 | Excessive mental workload | L1, L5 |
| H4 | Lack of explanation by ADS | L2, L3, L5 |
| H5 | Misjudgement by the ADS of the user's psychological state | L5 |
| H6 | Unexpected driving behaviour by ADS | L1, L2, L3 |
| H7 | ADS ignores human driver request | L1, L3, L3, L4, L5 |
| H8 | Lack of awareness by ADS (e.g. erroneous environment vehicle state data) | L1, L2, L4 |

### Specify psychological safety goals (psychological safety constraints)

Now that psychological losses and hazards are identified, systems constraints can be specified in order to prevent and/or mitigate the psychological losses and hazards. In our approach, this is done through psychological safety goals which we list in Table 4.

Table 4. Specified psychological safety goals

| ID | Psychological Safety Goal | Related Hazard |
|---|---|---|
| **PsySG1** | **Human driver must understand vehicle's automation capabilities & limitations** | **H1** |
| PsySG1.1 | Human driver must be able to understand the vehicle behaviour | |
| PsySG1.2 | Vehicle must ensure the human driver is familiar with the automation features capabilities & limitations (may be refined: e.g. reminders, quick tour of features, explanations) | |
| **PsySG2** | | **H2** |

| ID | Psychological Safety Goal | Related Hazard |
|---|---|---|
| PsySG2.1 | In the event of an unexpected situation (failures/external events), Vehicle must ensure the human driver is aware through clear messages/signals/visualizations | |
| PsySG2.2 | In the event of an unexpected situation (failures/external evens), human driver must be able to re-engage with the vehicle (take control/be aware of the situation) | |
| PsySG.2.3 | Ensure other passengers are aware through adequate measures (other signals to other passengers) | |
| **PsySG3** | **In the event of an unexpected situation, vehicle must remain in a supportive state for the human driver (perceived support)** | **H3** |
| | Vehicle must reassure the driver | |
| | Vehicle must facilitate the task to the human driver | |
| | Human driver must understand the vehicle autonomy capability and limitation | |
| | While handling unexpected situation, vehicle must keep the human driver updated on the evolution of the situation | |
| **PsySG4** | **Vehicle must ensure explainability of driving behaviour & messages** | **H4** |
| PsySG4.1 | Information feedback: vehicle must provide human driver with explanation for unexpected behaviours during & post events (implies the necessity of having a post event analysis capability.) | |
| PsySG4.2 | Anticipation: vehicle must anticipate the behaviour (detect malfunctions, design domain exit conditions), and communicate with the human driver [mitigates] | |
| PsySG4.3 | Comply with recommendations from existing standards on explainability | |

### 5.2.2. Build the control structure

The basic components of a STAMP control structure are made of a controlled process and one or more controllers which can be automated or human. Usually, a controller relies on a control algorithm, a model of the environment, and a model of the controlled process to generate control actions that are applied to the controlled process. Hence, typical components that make the model of a human controller in a STAMP control structure consist of mental models (i.e. human controller's beliefs about the environment, automated controller, and controlled process), mental model updates (how the human controller form their beliefs) and a decision-making process. Since humans don't function using fixed algorithms, the control algorithm is replaced by a decision-making process.

#### Model the control structure

The proposed control structure is presented on Figure 5**.** In our control structure, we were able to simplify the mental model updates by relying on Endsley's Situation Awareness (SA) model consisting of three levels (perception, comprehension and projection) (Endsley, 2000). We assert that these 3 levels can help determine how improper SA (e.g. improper perception or lack of comprehension) can lead to psychological losses. Although not common in a STAMP control structure, we added a psychological state component in the structure for capturing how negative psychological impact may occur. The ADS part of the control structure is based on SAE J316's functional architecture described earlier.

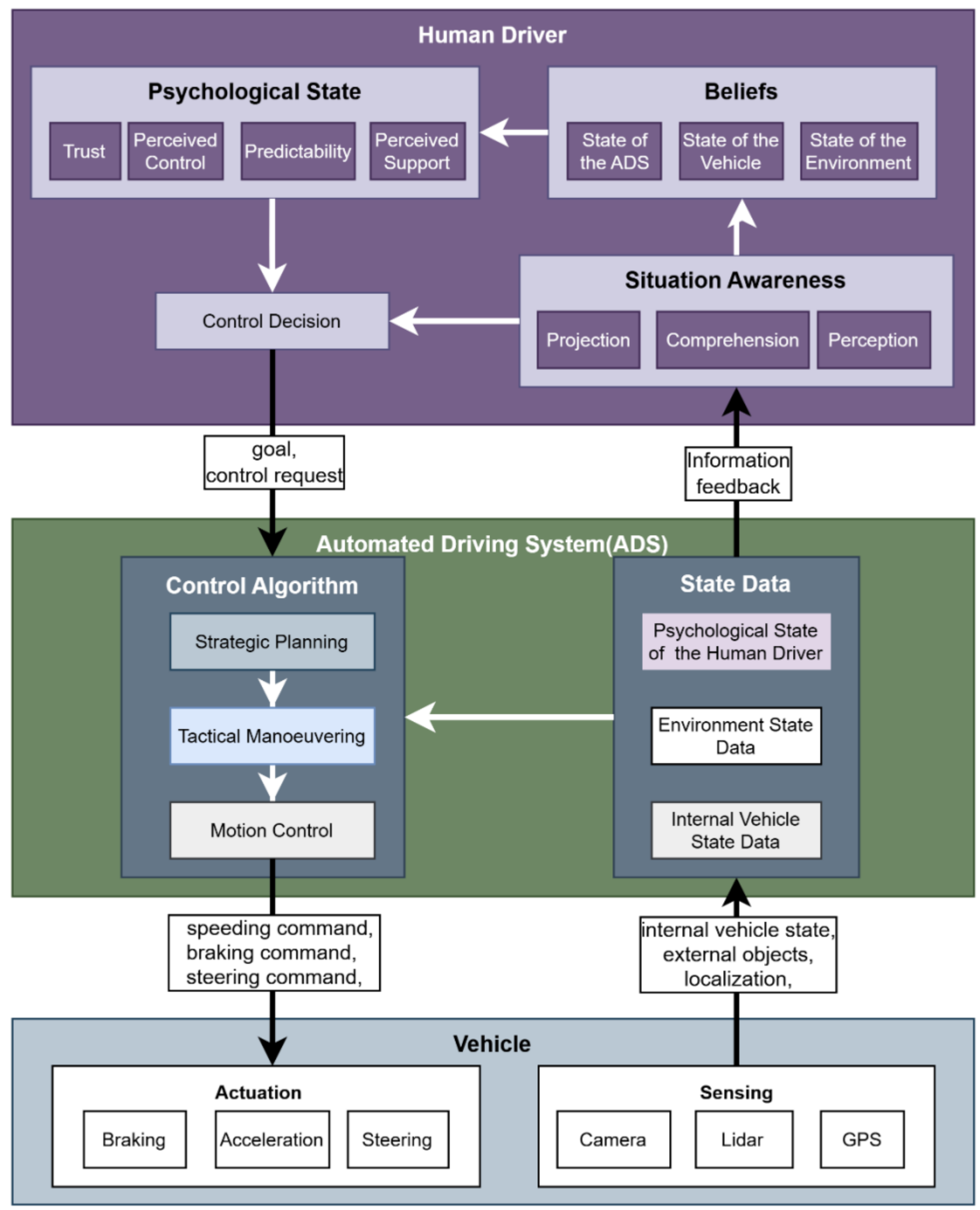

Figure 5. Psychological safety control structure

This control structure will serve in the responsibility assignment (i.e. the assignment of previously specified psychological safety goals to the system components). Additionally, it will also serve in the identification of the unsafe control actions and psychological loss scenarios in the third and fourth steps of the methodology.

## Assign Responsibilities

This sub-step assigns the measures previously specified through psychological safety goals (i.e. psychological system constraints) to relevant control structure components to ensure they are enforced. A responsibility is the rewriting or refinement of the derived psychological safety goal into a formal requirement.

Table 5. Responsibility assignment

| PsySG ID | PsySG Description | ID | Responsibility |
|---|---|---|---|
| PsySG1.1 | Human driver must have the ability to understand the vehicle behaviour (e.g. what the vehicle is doing and plans to do) | R1.1.1 | Human driver shall undergo awareness campaign to familiarize with AV |
| PsySG1.2 | Vehicle must ensure the human driver is familiar with the automation features capabilities & limitations (may be refined: e.g. reminders…quick tour of features, explanations…) | R1.2.1 | Vehicle must verify human drivers is familiar with automation features (awareness / training records etc.) and if not, vehicle shall be able to allow access to quick tour of features to ensure familiarity. |
| | | R1.2.3 | The vehicle shall be able to document that the driver had accessed the features descriptions. |
| | | R1.2.4 | The ADS shall be able to display its status (on/off/available/ in default/standby/in action) to the human driver (convey its capabilities through clear messages such as icon, sound, warning display) |
| | | R1.2.5 | The ADS shall inform the driver about the events that have been detected |
| | | R1.2.6 | The ADS shall inform the driver in case of actions of the capabilities that may psychologically impact the driver |
| PsySG2.1 | In the event of an unexpected situation (failures/external events), the vehicle must ensure the human driver is aware through clear messages/signals/visualizations | R2.1.1 | In case of immediate action performed by ADS, the ADS shall inform the driver in a specific manner to indicate the emergency |
| | | R2.1.2 | The ADS must ensure the quality of the message i.e. that the information it transmits is explainable (precise, succinct, clear, no room for interpretation) |
| PsySG2.2 | In the event of an unexpected situation (failures/external evens), human driver must be able to re-engage with the vehicle (take control/be aware of the situation) | R2.2.1 | The human driver shall expect requests from the ADS and in the case of an unexpected event, be receptive to messages and willing to re-engage with the vehicle |
| | | R2.2.2 | The ADS shall be able to inform the human driver to re-engage with the vehicle (to avoid or limit the surprise) |
| | | R2.2.1 | The human driver shall be in state to receive messages provided by ADS to re-engage with the vehicle |
| | | R2.2.2 | The ADS shall be able to inform the human driver to re-engage with the vehicle (to avoid or limit the surprise) |
| PsySG.2.3 | In the event of an unexpected situation, the ADS must ensure other passengers are | R2.3.1 | The ADS shall ensure that the messages are received by other passengers (to avoid or limit the surprise) |

| PsySG ID | PsySG Description | ID | Responsibility |
|---|---|---|---|
| | aware, ensure other adequate measures (to be defined) | | |
| PsySG.3 | In the case of an unexpected event, or behavior, vehicle must remain in a supportive state for the human driver (perceived support) | R.3.1 | In the case of an unexpected event, or behavior, ADS must be transparent in the decision leading to the performance of a specific dynamic driving task (tactical manoeuvring and motion control) |
| | | R.3.2 | Transparent about the contextual elements (internal & external) leading to a decision -> R.1.2.5 |
| | | R.3.3 | Vehicle must not be aggressive in the message in a manner not to shock or hurt |
| | | R.3.4 | ADS must remain a state of enjoyability (e.g. friendliness of messages) |
| PsySG.4.1 | Vehicle must ensure explicability of driving behaviour & messages | R.4.1 | ADS must provide feedback (what happened or is happening / intent or ongoing immediate action) |
| | | R.4.1 | Possess post event analysis capability (record events) |
| PsySG4.2 | vehicle must anticipate unexpected behavior of the ADS and communicate with the human driver for instance not to stress. | R.4.2 | ADS must detect malfunctions, design domain exit conditions that my lead to psychological impact or worsen psychological state |

### 5.2.3. Identify Unsafe Control Actions (UCA)

In the context of our example scenario, the following UCAs listed on Table 6 were identified. As it appears at the top of the table, the UCAs were identified following the four categories suggested by STPA as mentioned in previous section 4. The context of the UCAs is related to the use case scenario described earlier in section 5.2. To Identify the UCA, the four categories of unsafe control are applied to control actions from the previously proposed psychological safety control structure.

Table 6. Identified psychologically unsafe control actions

| **Control action (expected behaviour)** | **Providing** | **Not providing** | **too early, too late, or in the wrong order** | **lasts too long or is stopped too soon** |
|---|---|---|---|---|
| Provide braking command | **UCA1.1 -** ADS provides abrupt braking command during an emergency stop [H2, H6] | **UCA1.2** - ADS does not provides braking command during an emergency stop [H2, H6] | **UCA1.3** - ADS provides braking command too late during an emergency stop [H2, H6] | N/A |
| Inform human driver (by ADS) | **UCA2.1 -** ADS provides too much information [H2, H3] | **UCA2.2** - ADS does not provide explanation when performing an emergency braking [H2, H4] | **UCA2.3** - ADS Informs the Human driver too late when performing an emergency braking [H2, H6] | N/A |

| **Control action (expected behaviour)** | **Providing** | **Not providing** | **too early, too late, or in the wrong order** | **lasts too long or is stopped too soon** |
|---|---|---|---|---|
| Decide and request control (Human driver) | NA | **UCA3.2** Human driver does not request control (e.g. after ODD exit condition) | **UCA3.3** Human driver request control too late (e.g. into critical manoeuvre by ADS) | NA |
| Process feedback information (Human driver) | N/A | Human drivers ignore (does not process) feedback information | Human driver understands feedback information too late | N/A |
| Respond to human driver control request (ADS) | N/A | **UCA4.2** ADS does not respond to human driver's control request [H7] | **UCA4.3** ADS respond to human driver's control request too late [H6, H7] | N/A |

### 5.2.4. Identify psychological loss scenario

The identification of the loss scenarios is done by relying on the following four categories of causal factors explaining why a controller might provide (or not provide) a control action that is unsafe as provided by STAMP. These include failures involving the controller, inadequate control algorithm, unsafe control input, and inadequate process model. In the context of our illustrative scenario, the psychological loss scenarios listed in Table 7 were identified.

Table 7. Identified psychological loss scenarios

| **UCA** | **Control action** | **Scenario** | **PsySIL** |
|---|---|---|---|
| **UCA1.1** - ADS provides abrupt braking command during an emergency stop [H2, H6] | Provide braking command | **UCA1.1.SC1 (failures involving the controller) -** During an emergency stop, a failure (e.g. ABS) in the ADS leads to the braking command being applied too quickly/strongly?<br><br>**Inadequate control algorithm -** Erroneous interpretation of an operational situation (pedestrian & dog crossing) leading to inadequate control decision (braking)<br><br>**Unsafe control input -** During an emergency stop, the command is applied inadequately leading to unexpected /abrupt stop<br><br>**Inadequate process model -** | **C** |
| **UCA2.2** - ADS does not provide explanation when performing an | Inform human driver (by ADS) | **UCA2.2.SC1 (inadequate control algorithm)** | **B** |

| UCA | Control action | Scenario | PsySIL |
|---|---|---|---|
| emergency braking. [H2, H4] | | **failures involving the controller** - when performing an emergency braking, a bug in the ADS prevents the visual/auditive feedback being displayed/actuated.<br><br>**inadequate control algorithm** – inability of the ADS to interpret the situation: the ADS cannot interpret/process the situation leading to lack explanation?<br><br>**unsafe control input -**<br><br>**inadequate process model -**<br><br>**UCA1.2.SC2 (inadequate process model) -** | |
| **UCA3.3** Human driver requests control too late (e.g. into critical manoeuvre by ADS) | Decide and request control (Human driver) | **failures involving the controller (human error/mistake) -**<br><br>**inadequate control algorithm (inadequate control decision)-**<br><br>**unsafe control input - NA**<br><br>**inadequate process model (inadequate believe about the state of the vehicle) -** | **C** |
| **UCA4.2** ADS does not respond to human driver's control request [H7] | Respond to human driver requests (SG2, SG3) | **UCA4.2 - SC1 (inadequate control input)** – The ADS controller does not receive the human driver's request or receive the human driver's request too late due to faulty input from driver or inadequate priority arbitration in the control algorithm, leading to the request being ignored or delayed.<br><br>**UCA4.2 - SC2 (inadequate control algorithm)** - The ADS controller received the human driver's request. But based on the operational situation (e.g. rain, speed, surrounding vehicles, traffic filtering etc), the ADS controller asserted that takeover transition wouldn't leave enough time for the human driver to handle the situation appropriately. Thus, the ADS controller decided to handle the situation before giving control to the human driver. | **B** |

As described, the application of the method leads to the identification of psychological loss scenarios. Alternatively, we propose that methods used for dysfunctional scenarios expected by 26262/SOTIF can be explored for their adequacy to identify psychological loss scenarios. Moreover, some of these psychological scenarios may be related to functional safety or SotIF scenarios. Therefore, they can be mutually reused or revisited.

After identifying psychological loss scenarios, one intent of our adaptation is to supplement the step with a risk assessment based on PsySIL. Combining the severity assigned to the losses to an estimation of the controllability (deduced from the effectiveness of control measures defined through the psychological goals and responsibility) and the probability of occurrence based on the elements of the operational context. Moreover, PsySIL are assigned to the relevant psychological safety goals, thus allowing their refinement.

# 6. DISCUSSION

## 6.1. Risk model

This paper proposed an extended risk model and hazard analysis method for AV psychological safety based on the STAMP and STPA. The extended risk model refines the psychological state into four components: trust, perceived control, predictability, and perceived support and adds traceability relations to STPA artifacts, ensuring comprehensive coverage of psychological risk elements in the analysis. In addition, the proposal hypothesizes that these components can be objectively measured, allowing for a more tangible determination of psychological state. The refined model enables a more pragmatic estimation of the severity of psychological losses, which is crucial for assigning Psychological Safety Integrity Levels (PsySIL).

## 6.2. Systems theoretic process analysis adaptation

The proposed method adopts the four steps of STPA to incorporate psychological dimensions providing a structured approach to psychological safety analysis in the Autonomous Vehicles context. The adaptation includes assigning Psychological Safety Integrity Levels (PsySIL) to define the stringency with which psychological safety goals must be applied. This provides a quantitative aspect to psychological safety assessment and can be useful for the prioritization of psychological safety concerns

An adapted methodology for analysing psychological safety, particularly in the context of autonomous vehicles based on the STPA approach but has been modified to incorporate psychological risk factors. A risk model and an adapted system theoretic hazard assessment method were proposed and applied to a use case scenario.

## 6.3. Use case application

A use case application of the extended risk model and STPA adaptation focused on a specific scenario involving a SAE Level 4 autonomous vehicle operating in an urban/inner-city environment was executed. It involved the vehicle making a turn with limited visibility and encountering a pedestrian with a dog, resulting in an abrupt emergency stop. The analysis identified psychological losses and hazards based on the violation of psychological states derived from the specified four key psychological safety components including trust, perceived control, predictability and perceived support. Subsequently, psychological safety goals to prevent hazards and mitigate psychological losses were specified and assigned to system components. Some key goals included ensuring the human driver understands the vehicle's automation capabilities and limitations and providing explicability of driving behaviour and messages. Finally, causal factors leading to psychological losses were assessed through the identification of unsafe control actions (UCAs) and psychological loss scenarios considering factors like automated controller failures or inadequate control algorithms.

## 6.4. Limitations

### 6.4.1. Complexity and interdependency of psychological components

The proposal structured psychological safety into four components (trust, perceived control, predictability, and perceived support); and by doing so provided a ordered approach for its analysis. However, human psychology is highly complex, and these components are likely interdependent. Therefore, it is acknowledged that by treating the four psychological components as distinct, the approach may not fully capture the complex interactions between these psychological elements. Consequently, more in-depth expertise in both systems engineering and human factors may be needed.

### 6.4.2. Limited scope of use case and validation.

To illustrate the application of the proposal, the paper relied on a specific urban scenario with a Level 4 AV. This narrow scope may not capture the full range of psychological safety concerns that could arise in different environments, AV levels, user demographics, or where an occupant is not an operator. Therefore, the validation of the proposal's broader applicability may require more diverse use cases.

## 6.5. Future work

Based on the discussed limitations, three main avenues for future work are proposed. They include 1) an iterative framework development with experts and AV developers, 2) real-word application case studies and 3) the generalization of the framework to other autonomous systems and its integration with existing safety frameworks.

### 6.5.1. Iterative framework development with a panel of experts

The framework should be further evaluated with a panel of experts (including AV developers and psychology etc.), to gather feedback and use their input to refine the psychological safety components, risk model, and STPA adaptation. This would provide a validation of the proposed psychological safety risk model and hazard analysis method.

### 6.5.2. Various case study applications

Another avenue for future work could consist of working with AV development teams to apply the framework to real-world projects, using the framework in actual AV development scenarios (either past or in situ), to further refine the framework and demonstrate its practical applicability for real world use.

### 6.5.3. Generalization to other autonomous systems and integration with existing safety frameworks.

Finaly, future work could attempt to generalize this applicability to other autonomous systems or industries where psychological safety is also important. This can include applications in sectors such as aviation, remote drone operations, etc.

## 7. CONCLUSION

This extended psychological safety risk model and hazard analysis method for Autonomous Vehicles (AVs) based on STAMP and STPA represents a significant step forward in addressing the critical aspect of psychological safety in AV development. By refining the psychological state into four key components - trust, perceived control, predictability, and perceived support - and integrating these with STPA artefacts, the model provides a more comprehensive framework for analysing psychological risks in AVs. Through the adaptation of the STPA methodology to incorporate psychological dimensions, the proposal offers a structured approach to identifying and mitigating psychological hazards. Moreover, the introduction of Psychological Safety Integrity Levels (PsySIL) adds a quantitative aspect to the assessment, potentially allowing for more precise prioritization of safety concerns. The use case application demonstrated the ability of the proposed approach to identify specific psychological losses, hazards, and safety goals in a realistic AV scenario. This practical application highlights the potential of the approach to uncover psychological safety issues that might be overlooked.

However, the proposed has some limitations, including 1) the potential oversimplification of complex psychological factors, and 2) the focus on a narrow use case that may limit effectiveness of the validation of the proposal. Despite these limitations, the extended risk model represents a valuable contribution to the field of AV's psychological safety. It bridges an important gap between traditional safety engineering approaches and the psychological aspects of human-AV interaction. As AVs continue to develop and become more prevalent, such integrated approaches to safety analysis will be crucial in ensuring not just the functional safety of these vehicles, but also their psychological acceptability to users and the public.

Future work could focus on addressing the identified limitations, particularly in validating the framework with AV developers, expanding the proposal's applicability to diverse scenarios, integrating it with existing safety frameworks, which would enable the development of robust methods for measuring and validating psychological safety factors in real-world AV operations. In conclusion, while there is room for refinement and expansion, this extended psychological safety risk model and hazard analysis method represents a significant step towards more comprehensive and human-centered safety approaches in AV development.

## REFERENCES

Abror, A., & Patrisia, D. (2020). Psychological Safety and Organisational Performance: A Systematic Literature Review. *International Journal of Advanced Science and Technology*, *29*(5), 3634–3644. https://www.researchgate.net/publication/340998351

Baker, A. L., Phillips, E. K., Ullman, D., & Keebler, J. R. (2018). Toward an understanding of trust repair in human-robot interaction: Current research and future directions. *ACM Transactions on Interactive Intelligent Systems (TiiS)*, *8*(4), 1-30.

Cao, J., Lin, L., Zhang, J., Zhang, L., Wang, Y., & Wang, J. (2021). The development and validation of the perceived safety of intelligent connected vehicles scale. *Accident Analysis & Prevention*, *154*, 106092. https://doi.org/10.1016/J.AAP.2021.106092

DoD. (2012). *MIL-STD-882E*. https://assist.dla.mil.

Edmondson, A. C., & Lei, Z. (2014). Psychological Safety: The History, Renaissance, and Future of an Interpersonal Construct. *Annual Review of Organizational Psychology and Organizational Behavior*, *1*, 23–43. https://doi.org/10.1146/ANNUREV-ORGPSYCH-031413-091305

Endsley, M. R. (2000). *Situation Awareness Analysis and Measurement*.

Energy Institute. (2020). *Guidance on Human Factors Safety Critical Task Analysis.* 1–82.

*ISO 21448:2022 - Road vehicles — Safety of the intended functionality*. (n.d.). Retrieved March 5, 2024, from https://www.iso.org/standard/77490.html#lifecycle

*ISO 26262 [2018] Road vehicles-Functional safety-Part 1*. (2018). www.iso.org

Koopman, P., Ferrell, U., Fratrik, F., & Wagner, M. (2019). *A Safety Standard Approach for Fully Autonomous Vehicles*.

Lasota, P. A., Fong, T., Shah, J. A., & -Delft, B. (2017). A Survey of Methods for Safe Human-Robot Interaction. *Foundations and Trends® in Robotics*, *5*(4), 261–349. https://doi.org/10.1561/2300000052

Leveson, N. (2020). *Safety III: A Systems Approach to Safety and Resilience Safety III: A Systems Approach to Safety and Resilience Contents*.

Leveson, N. G. (2012). Engineering a Safer World: Systems Thinking Applied to Safety. *Engineering a Safer World*. https://doi.org/10.7551/MITPRESS/8179.001.0001

Leveson, N. G. (2017). Rasmussen's legacy: A paradigm change in engineering for safety. *Applied Ergonomics*, *59*(Pt B), 581–591. https://doi.org/10.1016/J.APERGO.2016.01.015

Leveson, N., & Thomas, J. (2018). *STPA Handbook*. https://psas.scripts.mit.edu/home/get_file.php?name=STPA_handbook.pdf

Madsen, M., & Gregor, S. (2000, December). Measuring human-computer trust. In *11th australasian conference on information systems* (Vol. 53, pp. 6-8).

*MIL-STD-882 D SYSTEM SAFETY*. (2000).

Moody, J., Bailey, N., & Zhao, J. (2020). Public perceptions of autonomous vehicle safety: An international comparison. *Safety Science*, *121*, 634–650. https://doi.org/10.1016/J.SSCI.2019.07.022

NASA Chang, T. (2006). *Human Reliability Analysis Methods Selection Guidance for NASA*.

Nordhoff, S., Stapel, J., He, X., Gentner, A., & Happee, R. (2021). Perceived safety and trust in SAE Level 2 partially automated cars: Results from an online questionnaire. *PLOS ONE*, *16*(12), e0260953. https://doi.org/10.1371/JOURNAL.PONE.0260953

Othman, K., Othman, & Kareem. (2023). Exploring the evolution of public acceptance towards autonomous vehicles with the level of knowledge. *InnIS*, *8*(8), 208. https://doi.org/10.1007/S41062-023-01180-Z

Prasetio, E. A., & Nurliyana, C. (2023). Evaluating perceived safety of autonomous vehicle: The influence of privacy and cybersecurity to cognitive and emotional safety. *IATSS Research*, *47*(2), 160–170. https://doi.org/10.1016/J.IATSSR.2023.06.001

Rubagotti, M., Tusseyeva, I., Baltabayeva, S., Summers, D., & Sandygulova, A. (2022). Perceived safety in physical human–robot interaction—A survey. *Robotics and Autonomous Systems*, *151*, 104047. https://doi.org/10.1016/J.ROBOT.2022.104047

SAE MOBILUS. (2021). *Taxonomy and Definitions for Terms Related to Driving Automation Systems for On-Road Motor Vehicles*. SAE International. https://doi.org/10.4271/J3016_202104

Schaefer, K. E. (2016). Measuring trust in human robot interactions: Development of the "trust perception scale-HRI". In *Robust intelligence and trust in autonomous systems* (pp. 191-218). Boston, MA: Springer US.

Shariff, A., Bonnefon, J. F., & Rahwan, I. (2017). Psychological roadblocks to the adoption of self-driving vehicles. *Nature Human Behaviour*, *1*(10), 694–696. https://doi.org/10.1038/S41562-017-0202-6

Sirgabsou, Y., Hardin, B., Leblanc, F., Raili, E., Salvini, P., Jackson, D., Jirotka, M., & Kunze, L. (2024). *A risk model and analysis method for the psychological safety of human and autonomous vehicles interaction*. https://arxiv.org/pdf/2411.05732

Su, B., Jung, S. H., Lu, L., Wang, H., Qing, L., & Xu, X. (2024). Exploring the impact of human-robot interaction on workers' mental stress in collaborative assembly tasks. *Applied Ergonomics*, *116*, 104224. https://doi.org/10.1016/J.APERGO.2024.104224

Taibi, Y., Metzler, Y. A., Bellingrath, S., Neuhaus, C. A., & Müller, A. (2022). Applying risk matrices for assessing the risk of psychosocial hazards at work. *Frontiers in Public Health*, *10*, 965262. https://doi.org/10.3389/FPUBH.2022.965262/BIBTEX

Serban, A., Poll, E., & Visser, J. (2020). A Standard Driven Software Architecture for Fully Autonomous Vehicles. *Journal of Automotive Software Engineering*, *1*(1), 20–33. https://doi.org/10.2991/JASE.D.200212.001

Winwood, P. C., Peters, R., Peters, M., & Dollard, M. (2012). Further validation of the psychological injury risk indicator scale. *Journal of Occupational and Environmental Medicine*, *54*(4), 478–484. https://doi.org/10.1097/JOM.0b013e3182479f77

Xing, Y., Lv, C., Cao, D., & Hang, P. (2021). Toward human-vehicle collaboration: Review and perspectives on human-centered collaborative automated driving. *Transportation Research Part C: Emerging Technologies*, *128*, 103199. https://doi.org/10.1016/J.TRC.2021.103199

Yagoda, R. E., & Gillan, D. J. (2012). You want me to trust a ROBOT? The development of a human–robot interaction trust scale. *International Journal of Social Robotics*, *4*(3), 235-248.

You, S., Kim, J.-H., Lee, S., Kamat, V., & Robert, L. (2018). Enhancing Perceived Safety in Human–Robot Collaborative Construction Using Immersive Virtual Environments. *SSRN Electronic Journal*. https://doi.org/10.2139/SSRN.3260634

## ANNEX

**Table 8. Psychological components evaluation methods**

| Component | Subcomponent | Qualitative Evaluation Methods | Quantitative Evaluation Methods |
|---|---|---|---|
| **Trust** | | • Situational Trust Scale for Automated Driving (STS-AD) (Holthausen et al., 2020)<br>• Trust in Automated Systems (TiAS) scale (Jian et al., 2000)<br>• Trust in Automation (TiA) scale (Körber, 2019)<br>• Reviews by (Kohn et al., 2021) and (Brzowski and Nathan-Robots, 2019)<br>• Human Computer Trust Instrument (Madsen & Gregor, 2000)<br>• Trust Perception Scale-HRI (Schaefer, 2016)<br>• Human-Robot Interaction Trust Scale (Yagoda & Gillan, 2012) | Gaze behaviour (Hergeth et al., 2016)<br>HRV (Petersen et al., 2019)<br>Behavioural automation use (Akash et al., 2020) |
| | Perceived Risk | • Scales for Perceived Situational Risk (PSR) and Perceived Relational Risk (PRR) (Li et al., 2019)<br>• 5-point Likert of environmental factors (Ha et al., 2020; Stange et al., 2022) | Gaze behaviour (von Stüplnagel, 2020) |

| **Perceived Control** | | • Pearlin Mastery Scale (Pearlin & Schooler, 1978)<br>• Locus of Control Scale (Rotter, 1966; Bellem et al., 2018)<br>• 7-point Likert (Schneider et al., 2021) | |
|---|---|---|---|
| | Usability | • System Usability Scale (SUS) (Brooke, 1996)<br>• Differential Emotion Scale (DES) (Jonsson et al., 2005)<br>• Post-Study System Usability Questionnaire (PSSUQ) (Lewis, 1992)<br>• Usefulness-Satisfaction Scale (Lund, 2001)<br>• User Experience Questionnaire (UEQ) (Laugwitz et al., 2008) | |
| | Workload<br>Mental Workload | • NASA-TLX (Hart, 1986)<br>• Driving Activity Load Index (Pauzié, 2008) | Eye gaze (Guettas et al., 2019; Hecht et al., 2019)<br>Heart rate (Guettas et al., 2019)<br>Driving detection response tasks (Guettas et al., 2019) |
| | Explanations | Self-reported trust (Ayoub et al., 2021; Kim et al., 2023)<br>Usefulness (Shen et al., 2022)<br>Likert scale of adjectives about the system (Koo et al., 2014) | |
| **Predictability** | | | |
| | Shock/Surprise | 5-point Likert (Kulić & Croft, 2007) | Eye Gaze (Affectiva, n.d., Agrawal & Peeta, 2021) |
| | Anxiety | Robot Anxiety Scale (RAS) (Nomura et al., 2006)<br>6-item scale for situational anxiety (Tluczek et al., 2009) | HR, Galvanic Skin Response (Rubagotti et al., 2022)<br>Electromyogram (Kulic & Croft, 2007)<br>Clicker during AV ride |
| | Familiarity | 7-point Likert (Sun et al, 2020; Cao et al., 2021, von Stüplnagel, 2020) | Measuring how many times participant has seen environment |
| **Perceived Support** | | | |
| | Explanations | Explanation effect on trust and subsequently perceived support (Kim et al., 2023) | Time to take control, in-lane distance, monitoring ratio and frequency (Petersen et al., 2019) |